\documentstyle[twoside,fleqn,npb,epsfig]{article}
\newcommand{\AmS}{{\protect\the\textfont2
  A\kern-.1667em\lower.5ex\hbox{M}\kern-.125emS}}

\title{ Review of recent results in spin physics.} 

\author{R. Windmolders \address{Universit\'e de Mons-Hainaut,
         \\
        B-7000 Mons, Belgium \\
\vspace*{2mm}
\it{E-mail:Roland.Windmolders@cern.ch}
}}%

\begin{document}

\begin{abstract}
Recent results in polarized DIS are reviewed. Particular emphasis is placed
on new measurements of transverse and longitudinal asymmetries, on the tests of the
spin sum rules and on the analysis of the spin structure function $g_1$ in 
perturbative QCD at NLO. 
\\
\vspace*{5mm}
\\
\begin{center}
\it {\large { Talk given at the 7th International Workshop \\
on Deep Inelastic Scattering and QCD, \\
April 19-23, 1999,  Zeuthen, Germany.}}
\end{center}
\vspace*{5mm}
\end{abstract}

\maketitle

\section{INTRODUCTION}

The aim of this introductory talk is to summarize the recent results
in {\it polarized deep inelastic scattering.}  After a brief reminder
of experimental aspects specific to spin physics, we will review successively
the new results on the transverse spin structure function $g_2$, the evaluation
of the first moments $\Gamma_1$ for the proton and the neutron, the ongoing
tests of the Gerasimov-Drell-Hearn sum rule, the fits of the $Q^2$
evolution of $g_1$ in perturbative QCD and the semi-inclusive asymmetries of
charged hadrons. In general the data mentionned in this review have become
public during the last year either in  articles or in
preprints. More recent (and therefore unpublished) data will
be discussed in the spin session and summarized in the conclusions of
this workshop.

\section { GENERAL FEATURES OF POLARIZED DIS 
EXPERIMENTS}  

\subsection{Spin asymmetries}

In contrast with unpolarized experiments which measure cross sections,
polarized DIS experiments measure spin asymmetries
\begin{equation}
 A = \frac{\Delta \sigma}{ 2 {\overline \sigma}}
\end{equation}
where $\Delta \sigma$ and $\overline {\sigma}$ are respectively the spin
dependent and the spin averaged cross sections:
\begin{eqnarray*}
\Delta \sigma  =
 \sigma^{\uparrow \downarrow} - \sigma^{\uparrow \uparrow},~ 
\\
{\overline \sigma} =
(\sigma^{\uparrow \downarrow} + \sigma^{\uparrow \uparrow} )/2 
\end{eqnarray*}
(the arrows correspond to the respective orientations of the beam and target
particle spins).
\\
The physics asymmetry $A$ is equal to the measured one divided by the
beam and target polarizations and by a factor $f$ accounting for the
fraction of polarizable nucleons in the target:
\begin{equation}
\nonumber
 A = \frac{A_{meas}}{P_B P_T f}
\end{equation}
The uncertainties on the factors
 $P_B$, $P_T$ and $f$ are important sources of systematic error 
on the spin asymmetry $A$ and  on the spin structure functions.     
\\
The  characteristics of recent ($>$ 1990) spin experiments, including 
the average values of the beam
and target polarizations and the corresponding errors, are listed
in Table~\ref{tab:charac}.                                           
Accuracies of
2 to 3 $\%$ are presently achieved for $P_B$ and $P_T$.     
\\

\begin{table*}[htb]
\caption {Spin experiments 1990-99.}
\label{tab:charac}
\vspace{8mm}
\begin{tabular}{|c|c|c|c|c|c|c|}
\hline  \hline
 Exp. &  Beam &   Energy    &  $P_B$ & Target    & $ P_T$        &   Ref.\\
      &        &  (GeV)     &  &          &               &    \\
\hline
 SMC  & $\mu$ & 100-190    &$0.795\pm 2.4 \%$ &   $C_4H_9OH$ (p) & $0.86 \pm 3.2 \%$ & \cite{smc_a1}         \\
      &       &      &     &  $C_4D_9OD$ (d) & $     0.51 \pm 2.0 \%$&      \\
      &       &      &     &  $NH_3$  (p)    & $     0.89 \pm 2.7 \%$     &  \\
\hline 
 E142 & $e$    & 19-23-26 & $0.36 \pm 3.1 \%$  & $^3He$ (n) & $\sim 0.33 \pm 7.0 \%$ &   \cite{e142_tot}  \\ 
\hline
 E143 & $e$    & 10-16-29  & $\sim 0.80 \pm 2.5 \%$  & $^{15}NH_3$(p) & $\sim 0.70 \pm 2.5 \%$ &  \cite{e143_tot}    \\
      &    &     &            & $^{15}ND_3$(d) & $\sim 0.25 \pm 4.0 \%$ &       \\
\hline
 E154 & $e$    &  48.3 & $0.82 \pm 2.5 \%$   & $^3He$ (n)& $\sim 0.38 \pm 5.0 \%$ &   \cite{e154_1,e154_2,e154_3} \\
\hline
 E155 & $e$    &  48.3 & $0.81 \pm 2.5 \%$   & $^{15}NH_3$ (p)& $\sim0.70 \pm 2.5 \%$ & \cite{e155_1,e155_2,e155_pub}\\
      &        &       &    & $^{6}LiD$ (d)  & $\sim 0.22 \pm 4.0 \%$               &                \\
\hline
HERMES & $e$   &  27  & $0.40-0.65 \pm 3.0 \%$      & $^3He$ (n)     & $0.46 \pm 5.0 \%$     & \cite{hermes_1,hermes_2,hermes_3}\\ 
       &       &       &    & ${\overline H}$ (p) &  $0.88 \pm 4.0 \%$  &         \\      
\hline  \hline
\end{tabular}
\end{table*}

\begin{figure}[htb]
\mbox{
\epsfxsize=6cm
\epsffile{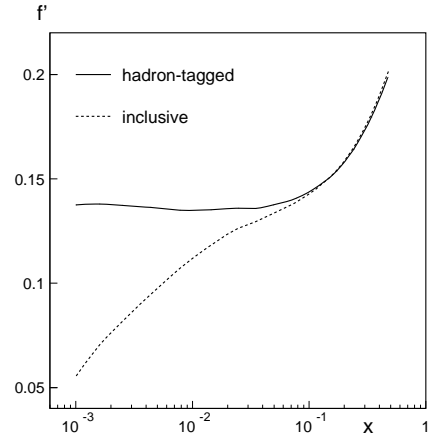}
}
\caption{ Effective dilution factor $f'$ for hadron tagged and inclusive events
from the SMC ammonia target.}
\label{fig:dilut}
\end{figure}

In its most simplified form the dilution factor $f$ is the ratio
of the number of polarizable nucleons in the target by the total number of nucleons
(e.g.$f = 3/17$ for 
$^{14}NH_3$).                     
More accurately, the numbers of nucleons of various types must be
weighted by the corresponding cross sections, e.g. in the
case of a proton target
\begin{equation}
f = \frac{n_p \sigma_{tot}^p}{\sum_A n_A \sigma_{tot}^A}
\end{equation}
with the sum in the denominator running over all target nuclei.
 \\
A significant dependence of the
dilution factor on the scaling variable $x$ is generated
by the variation of $F_2^n / F_2^p$ vs. $x$ as well as by  the
EMC effect modifying the cross section for nucleons bound in a nucleus. \\
In particular, the large radiative cross section at small $x$ in nuclei sharply reduces the 
value of $f$ for $x < 0.01$. In addition, a further decrease is produced by
radiative corrections to the cross section on the polarized nucleon itself.
\\
The measured asymmetry on a polarized proton target can be rewritten as
\begin{equation}
A_{meas} = P_B \cdot P_T \cdot f \frac{\Delta \sigma_{tot}^p}{\sigma_{tot}^p} =
 P_B \cdot P_T \cdot f \frac{\Delta \sigma_{tot}^p}{\sigma_{1 \gamma}^p} \cdot 
\frac{\sigma_{1 \gamma}^p}{\sigma_{tot}^p}  \\
\end{equation}
where $\sigma_{1 \gamma}^p$ is the one-photon exchange cross section and $\sigma_{tot}^p$
the total cross section including radiative effects.
Due to radiative corrections, the spin dependent cross section
$\Delta \sigma_{tot}^p$ slightly differs from 
$\Delta \sigma_{1 \gamma}^p$ and the previous formula becomes
\begin{equation}
A_{meas} = \Bigl ( \frac{\Delta \sigma_{1 \gamma}^p}{\sigma_{1 \gamma}^p} + RC \Bigr
 )P_B \cdot P_T \cdot f \cdot \Bigl ( \frac{\sigma_{1 \gamma}^p}{\sigma _{tot}^p} \Bigr )
\end{equation}
showing that the "effective dilution factor" is
\begin{equation}
f' = f \cdot \frac{\sigma_{1 \gamma}^p}{\sigma_{tot}^p}.
\end{equation}
 In the case of the SMC ammonia target, $f'$ drops to about 
 $ 0.05$ at $x = 0.001$.
\\
The effective dilution factor
can however be significantly enhanced by restricting the sample to events where 
a hadron is detected in the final state
 \cite{smc_a1}.
 In this case, the elastic tail does not contribute to the radiative corrections.   
The resulting  increase  of $f'$ at low $x$  largely compensates
the reduction of the number of events in the data sample so that more accurate values of the
asymmetry $A$ are obtained. For the SMC kinematics, this condition was
found to be true for $x < 0.02$. As an exemple, the $x$ dependences of $f'$ for inclusive and
hadron tagged events in the SMC ammonia target are shown in Fig.~\ref{fig:dilut}. With hadron
tagging, $f'$ is of the order of 0.14 and  approximatively constant at low $x$.

 \subsection {Virtual photon asymmetries and spin structure functions.}
The spin asymmetry (1) is related to the virtual photon asymmetries $A_1$ and $A_2$ by the 
following relations which refer to configurations where the beam and target
polarizations are parallel or perpendicular :
\begin{eqnarray}
\nonumber
A_{\parallel} = \frac{d \sigma^{\uparrow \downarrow} - d \sigma^{\uparrow \uparrow}} 
{d \sigma^{\uparrow \downarrow} + d \sigma^{\uparrow \uparrow}} = D (A_1 + \eta A_2), \\
\nonumber
\\
\nonumber
A_{\perp} = \frac{d \sigma^{\uparrow \rightarrow} - d \sigma^{\uparrow \leftarrow}} 
{d \sigma^{\uparrow \rightarrow} +d  \sigma^{\uparrow \leftarrow}} = D \lambda (A_2 + \eta ' A_1).
\end{eqnarray}
The depolarization factor of the virtual photon
\begin{equation}
D = \frac{y(2-y)}{y^2 + 2 (1-y)(1+R)}
\end{equation} depends on the ratio of the cross sections for longitudinally
and transversely polarized photons $R = \sigma_L / \sigma_T $ while the other
factors $(\lambda, \eta, \eta ')$ only depend on the event kinematics. The 
values of $A_1$ and $A_2$ are subject to the
positivity conditions:
\begin{eqnarray}
\nonumber
 |A_1| < 1, \hspace*{5mm} |A_2| < \sqrt{R}.
\end{eqnarray}
The spin structure functions $g_1$ and $g_2$ are 
obtained from the virtual
photon asymmetries $A_1$, $A_2$ and from the spin independent structure function $F_1$ :
\begin{eqnarray}
\nonumber
A_1&= (g_1 - \gamma^2 g_2)/F_1 ,\\ 
\nonumber
A_2&= \gamma(g_1 + g_2) / F_1
\end{eqnarray} 
with $ \gamma^2 = 4 M^2 x^2/Q^2$.\\
When $A_{\parallel}$ and $A_{\perp}$ are both measured, as in the SLAC experiments,
these relations fully determine $g_1$ and $g_2$. If only $A_{\parallel}$ is
measured, $g_1$ can be obtained from the relation
\begin{equation}
g_1 = \frac{1}{1 + \gamma^2} \Bigl [ \frac{A_{\parallel}}{D}+ (\gamma -\eta) A_2 \Bigr ] \frac{F_2}
{2 x (1 + R) }.
\end{equation}
In the kinematic range of the SMC experiment, 
the factors $\gamma$ and $\eta$ are both small, so that the contribution from the $A_2$ term
can be safely neglected. At lower energies (as in the HERMES experiment), the previous formula is used with a 
parametrization of $A_2$. 
\\
The uncertainties on the factors $P_B$,$P_T$ and $f$ induce a normalisation error of
about $4 \%$ on all values of $A_1$ and $g_1$. In addition,$A_1$ is affected by the
uncertainty on $(1+R)$, which varies with the kinematic variables and becomes important
in regions where $R$ is poorly constrained. This uncertainty partially cancels out in $g_1$ 
but the spin structure function is further affected by the uncertainty on $F_2$ which is of
the order of $4 \%$ (including a normalisation error of about $2 \%$).
\section{TRANSVERSE ASYMMETRIES}

\begin{figure}[htb]
\mbox{
\epsfxsize=6cm
\epsffile{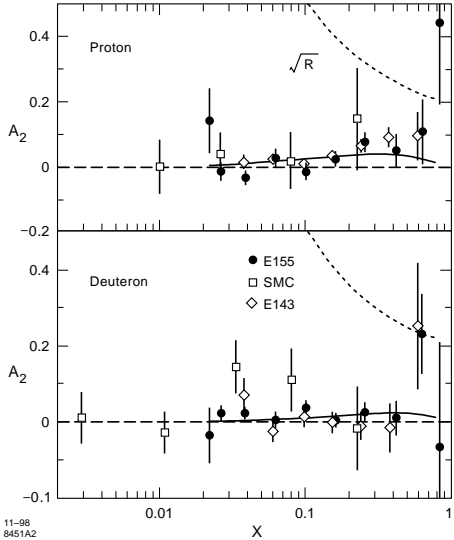}
}
\caption{ The asymmetries $A_2$ for proton and deuteron with their
statistical errors \cite{e155_1}. The solid lines show the twist-2 contributions
(corresponding to the WW term in $g_2$);
 the dashed lines show the positivity limits ${\sqrt R}$.}
\label{fig:a2}
\end{figure}

\begin{figure}[htb]
\mbox{
\epsfxsize=6cm
\epsffile{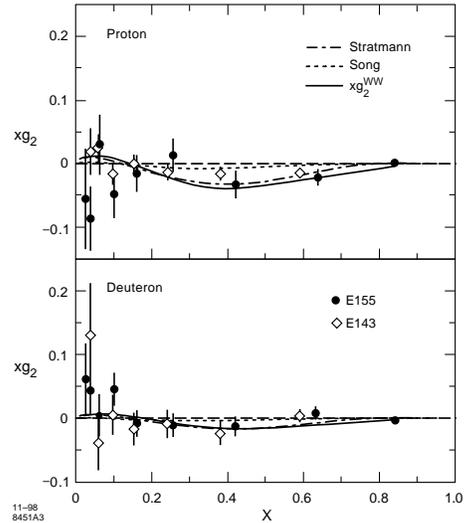}
}
\caption{The structure function $x g_2(x)$ from the experiments
E143 and E155 \cite{e155_1}. The errors are statistical. The full line shows
the twist-2 contribution $g_2^{WW}$; the dashed and dash-dotted lines show
model predictions  from ref. \cite{song} and \cite{stratman}.}
\label{fig:g2}
\end{figure}

Exploratory measurements of $A_2$ have been performed by the SMC in 1994-1996 for the
proton and deuteron targets. The transverse asymmetries were found to be
much smaller than their positivity  limit and compatible with zero within large
errors \cite{smc_perp_p,smc_perp_d}. $A_2^n$ was first measured by E142 and also
found compatible with zero \cite{e142_tot}. \\
The recent and more precise results from E143 \cite{e143_tot} and E155 \cite{e155_1}
show that $A_2^p$ is positive and significantly different from zero in the range
$0.2 < x < 0.7$ (Fig.~\ref{fig:a2}):
\begin{eqnarray*}
   < A_2>  & = 0.031 \pm 0.007    &(p), \\
   < A_2>  & = 0.003 \pm 0.013    &(d), \\
   < A_2>  & = -0.03 \pm 0.03     &(n).
\end{eqnarray*}
The structure function  $g_2$  is of special interest because it provides
a direct 
measurement of twist-3 contributions. It can indeed be decomposed as
\begin{equation}
g_2(x,Q^2) = g_2^{WW} (x,Q^2) + {\overline g_2}(x,Q^2)
\end{equation}
where
  $g_2^{WW}$ is the Wandzura-Wilczek term linear in $g_1$ and
 ${\overline g_2}$ an (almost) pure twist-3 contribution,
except for a negligeable contribution originating from transverse parton polarization.
The measured values of $g_2$ are in very good agreement with the Wandzura-Wilczek
term (Fig.~\ref{fig:g2}). As a consequence, 
the twist-3 matrix elements
\begin{equation}
d_n = 2 \int_0^1 x^n \Bigl ( \frac{n+1}{n} \Bigr ) {\overline g_2}(x,Q^2) dx
\end{equation}
derived from these data are consistent with zero.
At $Q^2$ = 5 GeV$^2$, for the combined SLAC data their values are:
\begin{eqnarray*}
   d_2^p   & = 0.007 \pm 0.004   \\
   d_2^n   & = 0.004 \pm 0.010.  
\end{eqnarray*}
 \\
\\
The Burkhardt-Cottingham sum rule (BC):
\begin{equation}
\int _0 ^1 g_2(x) dx = 0
\end{equation}
is  difficult to test on the existing data because the 
low $x$ behaviour of $g_2$ is unknown.
At $Q^2$ = 5 GeV$^2$ and for the range of the measurements
($0.02 < x< 0.8$), the combined SLAC data give values of              
$-0.015 \pm 0.026$ and 
$-0.010 \pm 0.039$ for the proton and the deuteron respectively,
in agreement with the BC  prediction. 
\\
The Efremov--Leader-Teryaev sum rule (ELT) 
predicts that the valence quark contribution to 
the second moment of ($g_1 + 2~g_2$)
must be zero :
\begin{equation}
\int _0 ^1 x (g_1^V (x) + 2 g_2^V (x)) dx = 0.  
\end{equation}
Assuming I-spin symmetry for the sea, this prediction becomes
\begin{equation}
\int_0^1 x \Bigl [ g_1 + 2 g_2 \Bigr ]^{p-n} dx = 0
\end{equation}
and is easier to test than the BC prediction because it
is less sensitive to the unmeasured contribution at low $x$ . At the same $Q^2$ and for the same
$x$ range, the SLAC data give 
$ 0.003 \pm 0.022$ , in good agreement with the ELT prediction.
 More accurate results on $g_2$ are expected in the near future from experiment E155X
which is finishing its data taking at SLAC. These new data will constraint
more precisely the values of the twist-3 matrix elements and may provide a discrimination
between a large number of models which are presently all compatible with the data
\cite{e155_1}.

\section {$\Gamma_1$ AND THE GDH SUM RULE}
\subsection {The first moment of $g_1$}
\begin{figure}[htb]
\mbox{
\epsfxsize=6cm
\epsffile{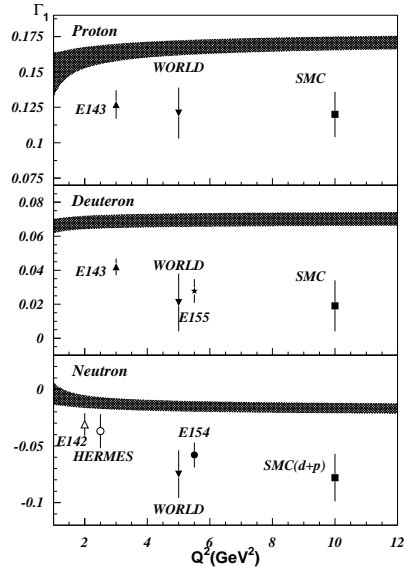}
}
\caption{Values of the first moment $\Gamma_1$ of the spin structure function
$g_1$ for the proton, deuteron and neutron. The shaded areas correspond to the
naive expectations assuming no contribution from strange quarks.}
\label{fig:ej_sum}
\end{figure}

The first moment of $g_1(x)$
\begin{equation}
\Gamma_1 = \int_0^1 g_1(x) dx
\end{equation}
is by far the most intensively discussed topic in polarized DIS.
The observation, made more than 10 years ago \cite{emc}, that the value
of $\Gamma_1^p$ is significantly smaller than the one expected from the
naive quark-parton model, assuming that the strange quarks do not contribute \cite{ej},
has been at the origin of a new generation of experiments. The initial
discovery has been confirmed by several new measurements, on the proton
and on the neutron, and over a wide range of $Q^2$ (Fig.~\ref{fig:ej_sum}).
\\
In the last years, special attention has been given to the low $x$ extrapolation
of the measured $g_1$ spectrum. At finite energy, $\Gamma_1$ is not a purely
experimental quantity and must be split as 
\begin{equation}
\Gamma_1 = \int_0 ^{x_{d}} g_1(x) dx + \int_{x_{d}}^{x_{u}}  g_1(x) dx
+ \int_{x_{u}} ^1 g_1(x) dx
\end{equation}
where the second term covers the measured range of $x$. While the high $x$
contribution is limited by the positivity condition $|A_1| < 1$ and by the
small values of $F_1$, the contribution from the low $x$ region is more
difficult to evaluate and could, in principle, be large. In the past, it was
estimated by a Regge extrapolation of $g_1$ at some low value of  $Q^2$
(generally $ \sim 1$ GeV$^2$). In recent analyses, this procedure is  replaced by
an extrapolation of the  shape of $g_1$ fitted at next-to-leading order (NLO) in QCD, which will be discussed
in a further section.
\\
This new approach results in a significant shift of $\Gamma_1$ towards
lower values
($\sim 0.015$ for the proton). It also shifts to a lower value  the singlet axial matrix element
$a_0$ which is derived from $\Gamma_1$ under the assumption that the non-singlet
part can be obtained from the $SU(3)$ coupling constants $F$ and $D$ as measured
in hyperon $\beta$ decay. For the proton and the deuteron, $a_0$ is related to $\Gamma_1$ 
by the formulas 
\begin{equation}
\Gamma_1^p = \frac{1}{18} ~C_{NS} ~(3 F + D) + \frac{1}{9} ~C_S ~a_0,
\end{equation}
\begin{equation}
\frac{\Gamma_1^d}{1 - 1.5 \omega_D} = \frac{1}{36} ~C_{NS} ~(3 F - D) + \frac{1}{9} ~C_S ~a_0
\end{equation}
where $C_S$ ans $C_{NS}$ are the singlet and non-singlet QCD coefficient functions
and $\omega_D$ is the probability for the deuteron to be in a D-state ($\sim 0.05$).
\\
The change in $\Gamma_1$ 
due to the low $x$ extrapolation reduces $a_0$ from the average value of 0.30 \cite{e143_tot}
to $0.18 \pm 0.09$ \cite{e155_pub}. In the naive QPM approach shown by the shaded
area in Fig.4, $a_0$ was expected to be identical to $a_8$ ($\simeq 0.58$).

\subsection {\bf The generalized GDH integral.} 
\begin{figure}[htb]
\mbox{
\epsfxsize=6cm
\epsffile{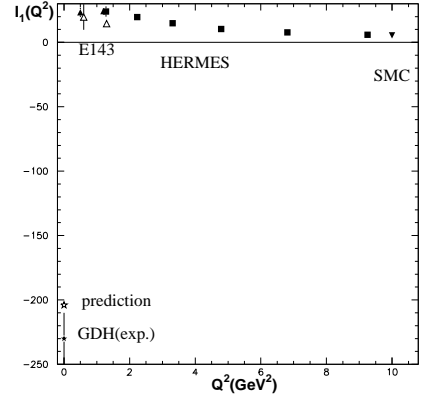}
}
\caption{Values of the GDH integral $I_1(Q^2)$  
(in $\mu b$) for the proton data from E143 \cite{e143_tot},
HERMES \cite{hermes_3} and  SMC \cite{smc_qcd} with the preliminary value
measured at $Q^2=0$ \cite{dhg_protvino} and the theoretical prediction.
In the E143 data,  the open
triangles correspond to the resonance region, the full triangles to the
DIS region.}
\label{fig:dhg}
\end{figure}

In general, spin structure functions depend on the variables $\nu$ and
$Q^2$. The function $g_1(x)$ discussed so far is a scaling limit
for large $\nu$ and $Q^2$:
\begin{equation} 
M^2 \nu G_1 (\nu,Q^2) \rightarrow g_1(x). 
\end{equation}
In order to study its behaviour at low $Q^2$ outside the scaling region,
the first moment $\Gamma_1$ should therefore be rewritten as
\begin{equation}
\Gamma_1 (Q^2) = Q^2 \int_{Q^2/2M} ^{\infty} \frac{M G_1(\nu,Q^2)}{2}
\frac {d   \nu}{\nu}.
\end{equation} 
The generalized Gerasimov-Drell-Hearn integral is defined by
\begin{equation}
I(Q^2) \simeq I_1(Q^2) = 16 \pi ^2 \alpha \frac {\Gamma _1 (Q^2)}{Q^2}
\end{equation}
and is of particular interest because its measured values (at finite $Q^2$) 
(Fig.~\ref{fig:dhg})
can be directly compared with the predictions of the GDH sum rule at $Q^2$ = 0 :
\begin{equation}
I(0) = \int _{\nu_0}^{\infty} \Delta \sigma (\nu) \frac{d \nu}{\nu} =
- \frac{2 \pi ^2 \alpha}{M^2} \kappa ^2
\end{equation}
where $\kappa$ is the anomalous magnetic moment of the nucleon. The predicted
values are
$- 204 ~\mu b$ for the proton and
$-233 ~\mu b$ for the neutron.
\\
Although it was derived more than 30 years ago \cite{gdh_th}, the GDH sum rule has
never been fully tested experimentally. A preliminary and partial result for the
proton covering the range $0.2 < E_{\gamma} < 0.8 $ GeV has been obtained last
year ($ - 230 \pm 20 ~\mu b$) \cite{dhg_protvino}. Combined with model calculations for
the missing contributions at low and high photon energy ($- 30$ and $+25 ~\mu b$ respectively),
it yields a value in agreement with the sum rule.
The experimental values of $I_1(Q^2)$ for the proton 
show a continuous
increase with decreasing $Q^2$ down to $Q^2 = 0.5$ GeV$^2$. A rapid variation
of the integral, including a change of sign, is thus required below this $Q^2$ 
if the data extrapolate smoothly to the sum rule prediction at $Q^2 = 0$.
As shown in Fig.~\ref{fig:dhg}, the integral $I_1$ remains positive in the resonance
region at the lowest  value of $Q^2$ measured by E143. However, negative asymmetries are
observed, as expected, in the $\Delta$ region \cite{e143_tot} and a rapid change 
of the integral in the resonance region is predicted at lower $Q^2$ \cite{dhg_mod}.
\\
A similar behaviour is expected for the neutron.
So far no measurements of the GDH sum rule have been made on a neutron target.
The generalized GDH sum is being evaluated
by experiment E94-010 at TJNAF using a polarized $^3He$ target  \cite{meziani_proc}.
The  data presently collected cover the range $0.03 < Q^2 < 1.0$ GeV$^2$ and $0.7 < W < 2.4$ GeV.   

\section {NEW DATA ON $A_1$ AND $g_1$.}
\begin{figure}[htb]
\mbox{
\epsfxsize=6cm
\epsffile{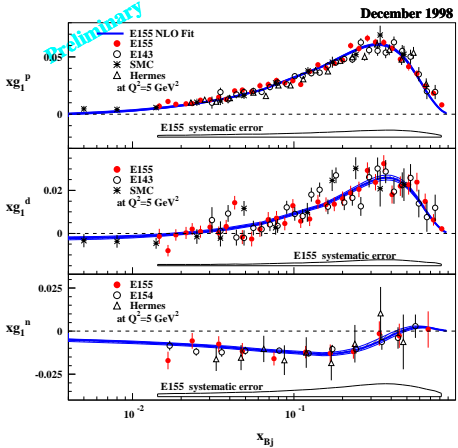}
}
\caption{ The structure functions $x g_1(x)$  at $Q^2 = 5$GeV$^2$ for the proton, the deuteron
and the neutron including preliminary results from E155 \cite{e155_pub}. 
(The lowest SMC point for $x g_1^p$ and $x g_1^d$ is not shown). The continuous lines show the
results of a QCD fit at NLO.}  
\label{fig:stackplot}
\end{figure}

\begin{figure}[htb]
\mbox{
\epsfxsize=6cm
\epsffile{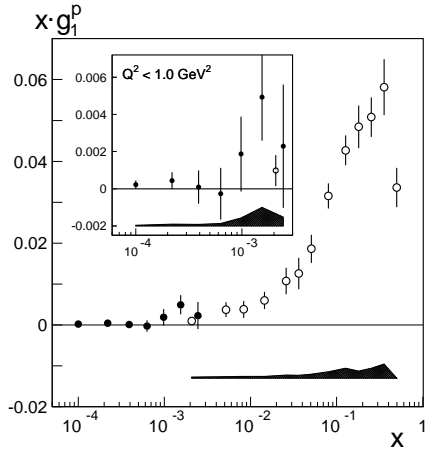}
}
\caption{ The values of $xg_1$ for the proton at the measured
$Q^2$ obtained with the SMC low $x$ trigger (filled circles) together with
those from the SMC standard triggers (open circles) \cite{smc_t15}.} 
\label{fig:smc_t15_p}
\end{figure}

\begin{figure}[htb]
\mbox{
\epsfxsize=6cm
\epsffile{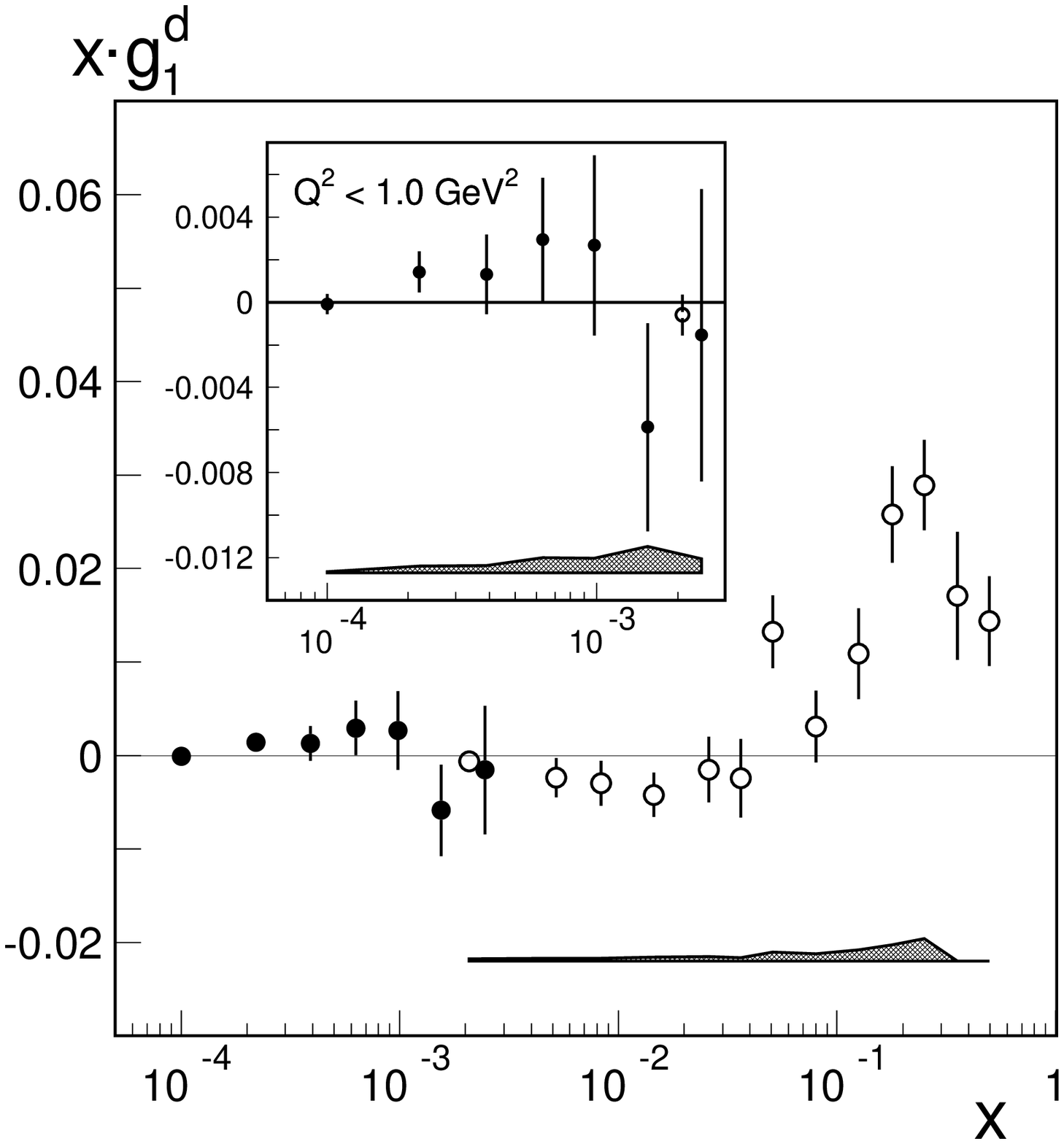}
}
\caption{ The values of $xg_1$ for  the deuteron at the measured
$Q^2$ obtained with the SMC low $x$ trigger (filled circles) together with
those from the SMC standard triggers (open circles) \cite{smc_t15}.} 
\label{fig:smc_t15_d}
\end{figure}

New data sets on the spin structure function $g_1$ have been published recently
by Hermes \cite{hermes_2} , SMC \cite{smc_t15} and E155 \cite{e155_pub}.
The deuteron data of E155 extend down to $x = 0.010$ and are in excellent agreement
with previous data from E143 and SMC when evolved to a common $Q^2$.
The Hermes data on hydrogen and the preliminary proton data from E155 also
agree remarkably well with previous data sets (Fig.~\ref{fig:stackplot}).
\\
The new SMC data on proton 
(Fig.~\ref{fig:smc_t15_p})
and deuteron 
(Fig.~\ref{fig:smc_t15_d})
cover a limited range at low $x$ and  low $Q^2$
($0.6 \cdot 10^{-4} < x < 0.8 \cdot 10^{-3} $ , 
 0.02 $< Q^2 < 0.2 $ GeV$^2$  ) 
which has never been explored so far. The use of a calorimeter signal in the trigger
and the requirement of a final state hadron in the analysis provide a clean sample
of $\mu N$ scatters in a kinematic region where $\mu~e$ scattering is the dominant
process.
  No significant spin effects are observed in the newly accessed region.
The values of $g_1$ shown in Figs.(7,8)  have been calculated with $F_2$ values taken from
the model of ref. \cite{ba_kw}.
The new data cover a very narrow range in $W$ ($14-16 $ GeV) and, for this reason, cannot be used to study the $x$
dependence of $g_1$  at fixed $Q^2$ in a Regge-type fit.

\section {QCD ANALYSIS OF SPIN STRUCTURE FUNCTIONS }
\subsection {Generalities}  
In QCD the quark-parton decomposition of the spin structure functions
 $g_1$   
\begin{eqnarray}
g_1^p (x) = \frac{1}{9} \Bigl [ 4 \Delta u + \Delta d + \Delta s \Bigr ],
\nonumber  \\
g_1^n (x) = \frac{1}{9} \Bigl [ \Delta u + 4 \Delta d + \Delta s \Bigr ].
\nonumber 
\end{eqnarray}
is replaced by the general relation 
\begin{eqnarray}
g_1^{p(n)}=\frac{1}{9} \Bigl ( C_{NS} \otimes \Bigl [ \stackrel{+}{-} \Delta q_{3} + \frac{1}{4} \Delta q_8
\Bigr ] 
\nonumber \\
+C_S \otimes \Delta \Sigma + 2N_f ~C_g \otimes \Delta g \Bigr ),
\end{eqnarray}
where 
$C_{NS},C_S,C_G$ are the non-singlet, singlet and gluon Wilson coefficients    and the symbol
$\otimes$ represents convolution with respect to $x$. In the case of 3 quark flavors discussed here,
the non-singlet ($\Delta q_3, \Delta q_8$) and singlet $(\Delta \Sigma)$ spin distributions are given in terms
of quark spin distributions by
\begin{eqnarray*}
 & \Delta q_3 = \Delta u - \Delta d, \\
 & \Delta q_8 = \Delta u + \Delta d - 2 \Delta s, \\
 & \Delta \Sigma = \Delta u + \Delta d + \Delta s.  
\end{eqnarray*}
The $Q^2$ evolution of the parton spin distributions is defined by the DGLAP equations
\begin{eqnarray}
\frac{d}{dt}\Delta q_{NS}
&=& \frac{\alpha_s(t)}{2\pi} P^{NS}_{qq} \otimes \Delta q_{NS},
  \\
\frac{d}{dt}\ \left (\matrix{\Delta \Sigma \cr \Delta g}\right)
&=&\frac{\alpha_s(t)}{2\pi}
\left (\matrix{P_{qq}^S & 2n_fP_{qg}^S \cr P_{gq}^S
& P_{gg}^S }\right) \otimes \left (\matrix{\Delta \Sigma \cr \Delta g}\right),
\nonumber
\end{eqnarray}
where 
 $t = \log{Q^2/\Lambda^2}$
and   
$P_{qq},P_{qg},P_{gq}$ are  polarized splitting functions.                                       
In this section, we will discuss fits of the $Q^2$ evolution at next-to-leading order
and review the results obtained in 4 different analyses performed by
Altarelli, Ball, Forte and Ridolfi ("ABFR",\cite{abfr}), Leader, Sidorov and Stamenov
("LSS",\cite{lss}), the E154 collaboration \cite{e154_3} and the SMC \cite{smc_qcd}.
\\
In all cases 
the procedure starts with  simple parametrizations of the polarized pdf's at some 
initial arbitrarely choosen $Q^2_i$. The pdf's are then evolved using the DGLAP equations (23) to the $Q^2$ of
every data point and the resulting $g_1$ recalculated. The  obtained values 
are compared to the measured ones and the parameters in the initial pdf's are
adjusted in order to minimize the resulting $\chi2$.
\\             
The results of the fit are used for several purposes. First of all, they provide a way
to evolve the measured values of
$g_1$ to a common $Q^2$ for the full range of the measurements.  
In the past, results were evolved to a common $Q^2$ under the assumption
that the ratio $g_1/F_1$ is independent of $Q^2$, a hypothesis which has no theoretical support
but is still compatible with the data for $Q^2 > 1$ GeV$^2$.                         
\\
The fitted shape of $g_1$ at fixed $Q^2$ is also used to extrapolate into the
regions out of the range of the experiments. At low $x$, the resulting 
shape is  quite different from the one obtained with a Regge
type extrapolation and leads to different estimates of $\Gamma_1$.
\\
Fits of $g_1$ are one of the few sources of information about 
polarized parton distributions. They are thus extremely useful to test or improve the
existing parametrizations.
\\
The  4 QCD  analyses discussed here are based on similar (although not identical) 
data sets and differ mainly by different choices concerning
the form of initial parametrizations, the definition of parameters and the initial $Q^2_i$.
The fitting procedure and the  handling
of systematic errors are also quite different.
\\ 
In the ABFR and SMC fits, the polarized pdf's are introduced explicitely,  in the form
\begin{equation}
\Delta q(x,Q^2_i) = \eta N(\alpha,\beta,a) x^{\alpha} (1 - x) ^{\beta} (1 + a x)
\end{equation}
where $\eta$, $\alpha$, $\beta$ and $a$ are free parameters
(some of them may be fixed in some cases). In the LSS and E154 fits,
the polarized pdf's are defined in relation with the unpolarized ones
\begin{equation}
\Delta q(x,Q^2_i) = \eta N(\alpha,\beta) x^{\alpha} (1-x)^{\beta} q(x,Q^2_i)
\end{equation}
which are taken from the MRST parametrization for LSS \cite{mrst}  or from the GRV parametrization
for E154 \cite{grv} of the valence and sea quark distributions. In this approach, a further assumption
has to be made, since inclusive spin asymmetries are not sensitive to valence and sea quarks.
\\
$\Delta {\overline s}$ can be obtained, at least in principle, from the difference
\begin{equation}
\Delta {\overline s} = (1/6) (\Delta \Sigma - \Delta q_8).
\end{equation}
Assuming
flavor symmetry breaking of the sea in the form $\Delta {\overline u} = \Delta {\overline d} = \lambda \Delta {\overline s}$,
one may then derive the valence quark spin distributions conditionnally on $\lambda$ :  
\begin{eqnarray}
\Delta u_v (\lambda) = \Delta u_v (\lambda=1) - 2 (\lambda - 1) \Delta \overline{s},  \nonumber  \\
\Delta d_v (\lambda) = \Delta d_v (\lambda=1) - 2 (\lambda - 1) \Delta \overline{s}.
\end{eqnarray}
The consistency of the results for different choices of  $\lambda$ with the above equations has been tested by LSS \cite{lss}.
It is worth mentionning that the introduction  of unpolarized pdf's (as in eqn.(25)) provides a satisfactory description
of the data 
with a smaller number of free parameters.
\\
Fits in the 4 analyses have been performed in moment space. The SMC analysis has used the same fitting
algorithm as the ABFR analysis but the results have been cross-checked with an independent program
where the fit is done in ($x-Q^2$) space.
\subsection { Factorization scheme}
At next-to-leading order, the splitting functions, the coefficient functions and, in general,
the parton distributions depend on the renormalization and factorization schemes, while
physical observables remain scheme independent.
In the  $\overline{MS}$ scheme, the gluon density does not contribute to the first moment $\Gamma_1$
because the  first moment of the gluon coefficient function is zero. In this scheme,
the singlet axial matrix element $a_0$ (eqns.(16-17)) is identical to the first moment of the singlet
distribution $\Delta \Sigma$ and depends on $Q^2$.
\\
The Adler-Bardeen scheme (AB) is a modified 
 $\overline{MS}$ scheme where $\Delta \Sigma$ is changed by terms proportionnal to 
 $ \alpha_s(Q^2) \Delta g(x,Q^2)$. These corrections 
  may be  large even at high $Q^2$  because $\int _0 ^1 \Delta g(x,Q^2) dx
\simeq \Bigl [ \alpha_s (Q^2) \Bigr ] ^{-1}$.
The singlet distributions in the AB and $\overline{MS}$ schemes are related by
\begin{eqnarray}
\Delta \Sigma (x,Q^2)_{AB} = \Delta \Sigma (x,Q^2)_{\overline{MS}} + 
 \nonumber \\ N_f
\frac{\alpha_s(Q^2)}{2 \pi} \int_x^1 \frac{dy}{y} \Delta g (y,Q^2)
\end{eqnarray}
while the gluon and non-singlet distributions remain unchanged. Consequently, the first moment
of the singlet distribution $\Delta \Sigma _{AB}$ differs from $a_0$:
\begin{equation}
a_0(Q^2) = \Delta \Sigma (1)_{AB} - N_f \frac{\alpha_s(Q^2)}{2 \pi} \Delta g(1,Q^2)
\end{equation} 
and is independent of $Q^2$. \\
It has been pointed out \cite{lss} that the AB scheme is
a particular case of a family of schemes differing by higher moments 
 $\Delta \Sigma (n)$, ($n \geq 2$).
\\
Fits in different schemes are expected to give consistent results, i.e. the different fitted  singlet
distributions should be in agreement with eqn.(28).

\subsection { Positivity conditions}
The condition that the difference of cross sections for total spin 1/2
and total spin 3/2 has to be smaller than the sum
\begin{equation}
| \sigma _{1/2} - \sigma _{3/2} | \leq \sigma_{1/2} + \sigma_{3/2}
\end{equation}
leads to the well known limits 
$|A_1| \leq 1 $ or
$|g_1(x,Q^2)| \leq F_1(x,Q^2)$. At leading order, these conditions imply
that
\begin{equation}
|\Delta q_i (x,Q^2)| \leq q_i(x,Q^2)
\end{equation}
 for all quark flavors. The same relation can be established for the gluon
by considering Higgs production by the process $g + g \rightarrow H$.
At next-to-leading order, these relations, which correspond to the
probabilistic interpretation of polarized pdf's, are no longer valid.
Their generalisation has been studied in ref.\cite{abfr} and  leads to 
correlated boundary conditions on the moments ($\Delta \Sigma(N),\Delta g(N)$). 
These conditions could be used as additionnal constraints in the determination of
polarized pdf's.

\subsection {Results on moments }
\begin{figure}[htb]
\mbox{
\epsfxsize=7cm
\epsffile{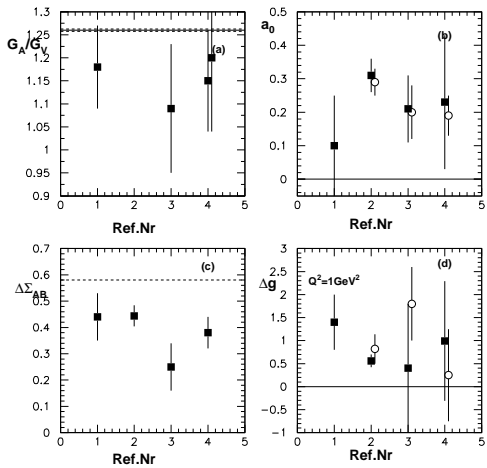}
}
\caption{(a) Values of $G_A/G_V$; (b) values of the singlet axial matrix element $a_0$;
(c) values of the singlet first moment in the AB scheme $\Delta \Sigma _{AB}$; (d) values of 
$\Delta g (Q^2=1$ GeV$^2$) in the 4 analyses (1=ABFR, 2=LSS, 3=E154, 4=SMC). 
Results obtained in the $\overline{MS}$ scheme are shown as open circles, results
obtained in the AB scheme as full squares. 
The quoted errors are the statistical and systematic errors combined in quadrature, however
the theoretical uncertainty is not included in the LSS errors. See text for the duplicated SMC point in part (a).}
\label{fig:bj_a0_dsig_gl}
\end{figure}

The results on the moments of the pdf's obtained
in the 4 different analyses are summarized in Fig.~\ref{fig:bj_a0_dsig_gl}.
\\
The first one,
 $\Delta q_3$, is often fixed in the fits at the nominal value of  (F+D) in order to satisfy
 the  Bjorken sum rule \cite{bj}. When it is left as a free parameter, the fitted value provides a
test of this sum rule. The results shown in the plot (part (a)) indicate that the data
confirm the Bjorken sum rule within an accuracy of about $10 \%$. The second result quoted
for the SMC analysis corresponds to a partial fit of the non-singlet part $g_1^p - g_1^n$.
Such a fit has the adventage that it requires only
few parameters and is independent of the gluon. The result is consistent
with the global fit but slightly less accurate, due to the limited amount of data presently usable in
this non-singlet fit.
\\
The values of the singlet axial matrix element $a_0$ (part (b) of the plot) are consistent in the 4 analyses.
They are also consistent 
 for fits done in the $\overline{MS}$ and AB schemes, as expected for a scheme independent
quantity. The errors are significantly larger for the derivations in the AB scheme, due to the additionnal
systematic error related to the gluon. It can also be seen that systematic errors have been treated
differently in the 4 analyses. The effect of changes in the factorization and renormalization scales by factors
varying
between 0.5 and 2.0 are included in the error quoted by the SMC analysis, as well as the effect of changes
in the form of the input parametrizations (24). By comparison, the errors quoted by the LSS analysis appear
largely underestimated.
\\
The first moment of the spin singlet in the AB scheme 
(part (c) of the plot)  
averages around 0.40. 
The comparison with the values of $a_0$ shows
the increase due to the gluon contribution to the nucleon spin. It is
however clear that the value of $\Delta \Sigma_{AB}$ remains significantly below the quark-parton model
expectation of 0.58. In other words, the present data suggest that the gluon accounts for about half of the
difference between the measured values of $\Gamma_1$ and their QPM expectations (Fig.~\ref{fig:ej_sum})
leaving room for other effects, such as orbital momentum.
\\
The first moment of the gluon spin distribution evaluated at $Q^2 = 1$ GeV$^2$ ( part (d) of the
plot) is of the order of 1. The same remarks as above apply for the errors. It is also  observed
that the evaluations made in the 2 factorization schemes differ by about one standard deviation but that
the sign of these differences is not the same in all analyses. This clearly shows the limit of attempts to
determine the gluon contribution from inclusive measurements where the role of the gluon is restricted
to the $Q^2$ evolution. 

\subsection {Results on parton spin distributions.}

Contrary to $\Delta q_3(x)$ which is unambiguously defined by the difference
$g_1^p(x) - g_1^n(x)$, the contribution to $g_1$ from $\Delta q_8(x)$, $\Delta \Sigma (x)$ and $\Delta g(x)$
can only be disentangled by their different $Q^2$ evolution.
The range in $Q^2$ is presently defined by the SLAC experiments ($E= 20$ GeV) and the SMC experiment
($E= 190$ GeV) and corresponds to a factor of about 6 at fixed $x$.
It is thus expected that only the largest 
contribution to $g_1$ will be reasonnably well constrained by the data.
\\

\begin{figure}[htb]
\mbox{
\epsfxsize=6cm
\epsffile{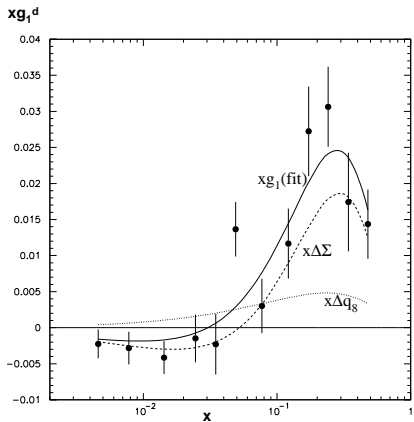}
}
\caption{The fitted $g_1^d(x)$  with the SMC data
at their measured $Q^2$; also shown are
the contributions from the singlet term (in the $\overline{MS}$ scheme)
and from the non-singlet term $\Delta q_8$.                 
}
\label{fig:deut_sing}
\end{figure}

Fig.~\ref{fig:deut_sing} shows the fitted $g_1^d(x)$ at the $Q^2$ of the SMC data together with
its non-singlet and singlet components. It can be seen that the singlet term $x \Delta \Sigma$
(evaluated in the $\overline{MS}$ scheme)
closely follows the variation of $x g_1^d$, from negative values at small $x$ to a maximum around 
$x = 0.25$. 

\begin{figure}[htb]
\mbox{
\epsfxsize=6cm
\epsffile{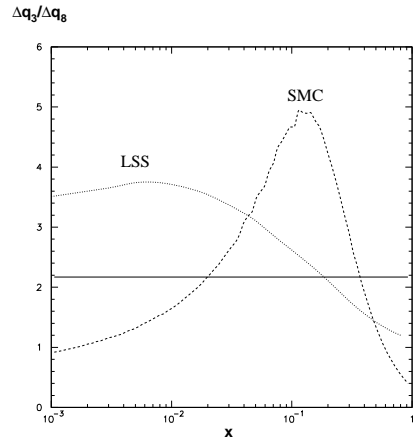}
}
\caption{The fitted ratio $\Delta q_3(x)/\Delta q_8(x)$ 
at $Q^2 = 1$GeV$^2$ as obtained in the LSS and SMC fits where the shape of
$\Delta q_8(x)$ is left free. The horizontal line shows the ratio of the 2 moments.                                       
}
\label{fig:q3_by_8}
\end{figure}

\begin{figure}[htb]
\mbox{
\epsfxsize=6cm
\epsffile{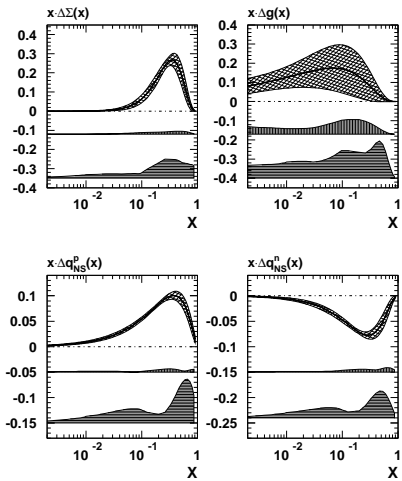}
}
\caption{Polarized parton distributions at $Q^2~=~1$GeV$^2$ from the SMC fit in the AB scheme
(singlet, gluon, non-singlet for the proton and for the neutron) \cite {smc_qcd}.
The bands with crossed hatch show the statistical uncertainty as obtained in the fit
while the vertically and horizontally hatched bands correspond to the experimental
and theoretical uncertainties respectively.
}
\label{fig:figure5}
\end{figure}

The singlet distribution appears thus to be strongly constrained by the measured values of $g_1^d$ 
and will be well determined by the fit. Results from different fits \cite{smc_qcd,lss}   
show indeed remarkable agreement. \\
The previous figure also suggest that the determination of the small $\Delta q_8(x)$ 
contribution will be much more difficult. In general, the first moment of $\Delta q_8$ is
fixed to the value obtained from hyperon $\beta $ decay (= 3F - D). In addition, its
shape is often assumed to be the same as $\Delta q_3(x)$. The latter assumption has no
theoretical justification and is motivated only by the lack of constraint in the fit.
As an illustration, we show in Fig.~\ref{fig:q3_by_8} the ratio $\Delta q_3/\Delta q_8$
obtained in the LSS and in the SMC fits where the
shape of $\Delta q_8(x)$ was left free. The deviations with respect to the ratio of 
moments (shown by the horizontal line) are completely different in the 2 fits and result mainly from the
constraints imposed by the analytical form of the input distributions. In conclusion, no
significant information on $\Delta q_8$ can be derived from the inclusive data with their present
precision.
\\
The same could be said about the shape of the gluon distribution which is affected by
very large statistical and systematic errors. 
Fig.~\ref{fig:figure5} shows the fitted pdf's in the AB scheme as obtained in the SMC analysis \cite{smc_qcd}. 
The statistical error  is much larger for the gluon than for the other distributions
because the gluon  only contributes to $g_1$ through the $Q^2$ evolution. For the same reason, the
gluon is strongly affected by the "theoretical" error originating from the variation in
renormalization and factorization scales.

\begin{figure}[htb]
\mbox{
\epsfxsize=6cm
\epsffile{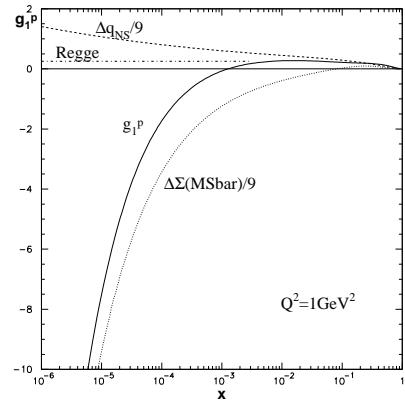}
}
\caption{The extrapolated $g_1^p(x)$ at $Q^2 = 1$ GeV$^2$ from the SMC QCD fit
at NLO compared to a Regge type extrapolation. Also shown are the extrapolated
singlet (in $\overline{MS}$ scheme) and  non-singlet terms.                 
}
\label{fig:lowx}
\end{figure}

As mentionned before, the fitted shape of $g_1(x)$ is extrapolated down to $x=0$ in order
to evaluate the moments $\Gamma_1$. Fig.~\ref{fig:lowx} shows the extrapolation of $g_1^p$ at
$Q^2 = 1 $GeV$^2$. The spin structure function becomes negative below $x = 0.001$ (i.e. slightly below
the lowest data point). The drop towards large negative values when $x$ tends to zero is
driven by the singlet term (also shown in the figure). It should be noticed that the polarized
pdf's at $x$ below the range of the data do not influence  the fit results (as shown
in eqn.(22), the value of $g_1$ at a given $x$ only depends on the pdf's at {\it higher} $x$). 
The validity of the extrapolation to $x = 0$ therefore rests on the assumption that the  shape
of polarized pdf's which was found to describe the data remains valid at lower $x$. Future
measurements of $g_1^p$ below $x = 0.001$, 
for instance from the polarized HERA collider \cite{pol_hera}, would be needed to confirm the change of sign suggested
by the extrapolation. 
\section {HADRON ASYMMETRIES }
\begin{figure}[htb]
\mbox{
\epsfxsize=6cm
\epsffile{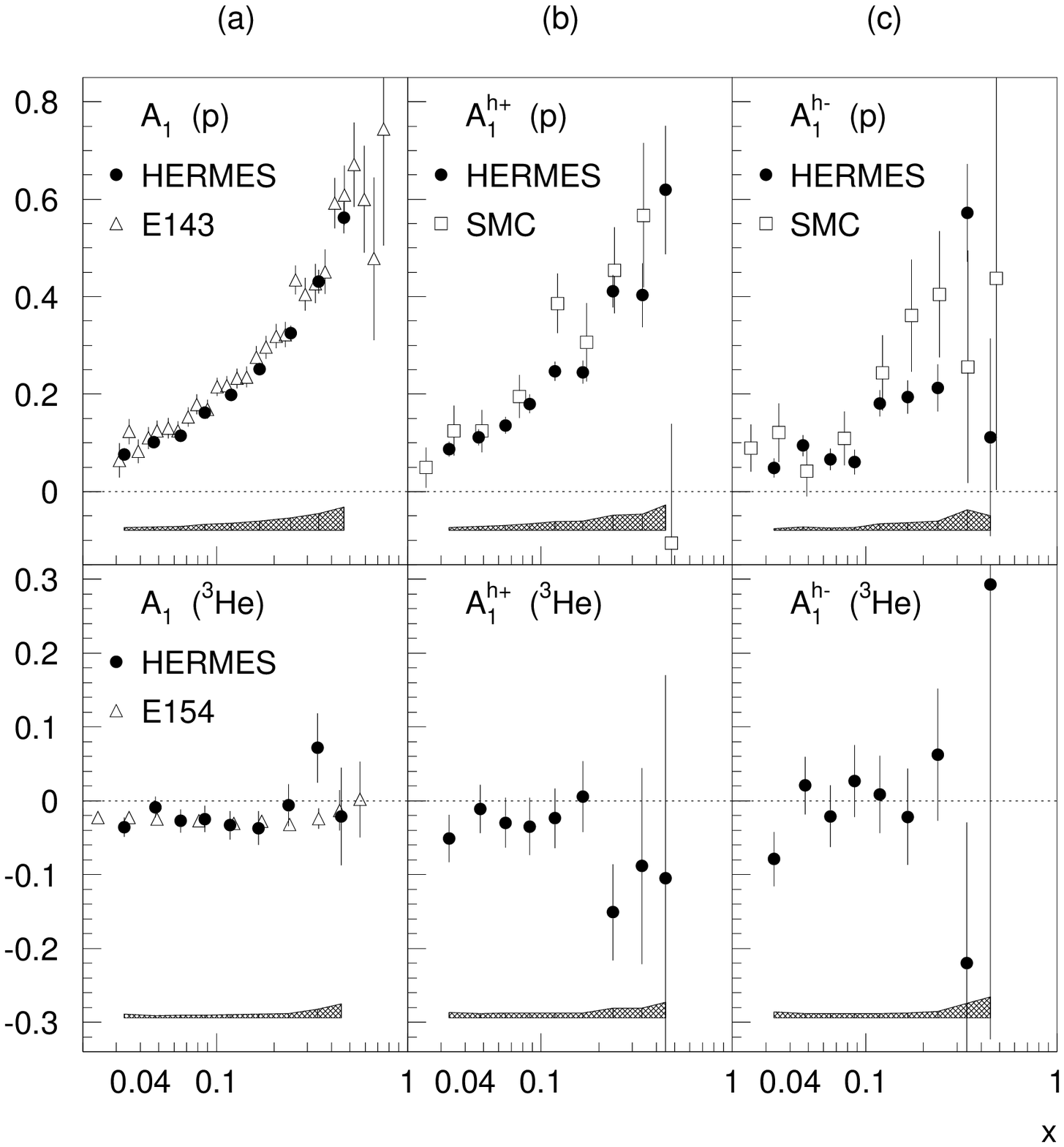}
}
\caption{Inclusive asymmetries (left side), semi-inclusive asymmetries 
for positive hadrons (center) and for negative hadrons (right side) measured
by HERMES. The upper and lower plots are for the hydrogen 
and $^3He$ target respectively. Data from previous experiments are shown for comparison.
}
\label{fig:hermes_asym}
\end{figure}

\begin{figure}[htb]
\mbox{
\epsfxsize=6cm
\epsffile{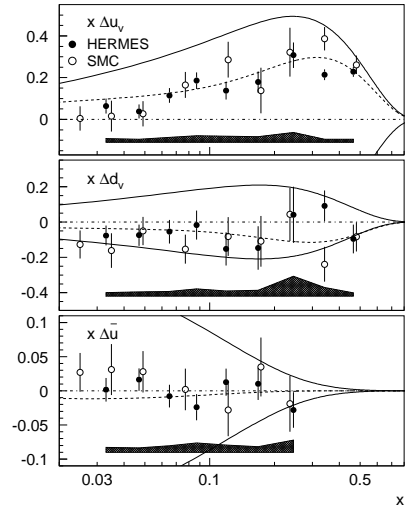}
}
\caption{Distribution of the valence and sea quark spin distributions derived
from the inclusive and semi-inclusive asymmetries measured by HERMES. The values
obtained previously by SMC in the same $x$ range are shown for comparison. The full
lines show the limits defined by spin independent quark distribution; the dashed lines are the prediction
of the G-S parametrization \cite{gs}.
}
\label{fig:hermes_si}
\end{figure}

Measurements of semi-inclusive asymmetries for positively and negatively
charged hadrons, when combined with the inclusive asymmetries and analyzed
in the framework of the quark-parton model, provide the spin distributions of valence
and sea quarks. The SMC analysis \cite{smc_si} has shown that the $u$ valence quarks
are polarized positively ($\Delta u_v = 0.77 \pm 0.13$) and 
that their polarization increases with $x$ while the $d$ valence quarks are 
polarized negatively ($\Delta d_v = -0.52 \pm 0.17$). No significant polarization
was found for sea quarks ($\Delta \overline{q} = 0.01 \pm 0.05$). \\
The semi-inclusive asymmetries collected by HERMES on the proton 
target are in good agreement with those from SMC and improve considerably the statistical
precision
(Fig.~\ref{fig:hermes_asym}).     
Combined with the less precise asymmetries on the $^3He$ target, these new data lead to the 
valence and sea quark spin distributions shown in 
Fig.~\ref{fig:hermes_si}. The SMC results are confirmed with statistical errors reduced by half
for the $u$ valence quark and the sea quarks. A comparable improvement is expected for the
$d$ valence quark when the data presently collected on a deuterium target will become available.
\\
In general, the analysis of semi-inclusive spin asymmetries has been limited to the
framework of the quark-parton model. A QCD analysis at next-to-leading order based on inclusive and
semi-inclusive data has been attempted in ref. \cite{deflo}. The emergence of more precise
semi-inclusive data, possibly also for identified particles, will make this approach more
relevant in the future and may help  answer some of the pending questions in the QCD analysis
of $g_1$.
\\
The difficulty to determine the gluon spin distribution from the $Q^2$ evolution of $g_1$  has
lead to an increased interest for reactions dominated by photon-gluon fusion, a process which
gives access to $\Delta g$ with a large analysis power. The COMPASS experiment, 
planned to start at CERN in the year 2000, intends
to determine $\Delta g$ from open charm production \cite{compass} using the full virtual photon flux
down to $Q^2 = 0$ and also from events where a pair of high $p_t$ hadrons is produced. 

The SLAC experiment E155 has studied the reactions
\begin{eqnarray*}
\gamma p \longrightarrow ({\rm hadron})^{\pm} ~X ,\\
\gamma d \longrightarrow ({\rm hadron})^{\pm} ~X~ 
\end{eqnarray*}
with circularly polarized photons \cite{e155_2}. In these reactions, the sensitivity
to the gluon spin distribution could also be enhanced due to
the photon-gluon process.                                           
Small asymmetries significantly different from zero have been observed 
for positive and negative hadrons produced on the proton target while the asymmetries on
the deuteron target are consistent with zero. The interpretation of these results remains difficult
because the event kinematics is unknown and many different processes may contribute.

\section {CONCLUSIONS}

The field of  spin physics has been characterized in 98-99 by 
a rather limited amount of  new data but  very extensive work 
 to  finalize the analysis of previous experiments.
Besides the ongoing analyses of spin structure functions in perturbative QCD, there has been	
a diversification of interest towards  the non-perturbative low $Q^2$ region and 
photoproduction, including the long awaited test of the Gerasimov-Drell-Hearn 
sum rule.
\\
 Hadron asymmetries have become a major subject of interest due to the precise data obtained
by HERMES.
\\
The transverse asymmetry $A_2$  is now well measured and found to be  different from zero for the proton.
Its compatibility with the twist-2 contribution will be further investigated with the most recent SLAC data.
\vspace*{3mm}
\\
It is a pleasure to thank my colleagues 
and friends from the SMC for many fruitful discussions over many years.
I also like to thank  colleagues from other collaborations who have kindly provided
information about their experiment during the preparation of this talk, in particular
A.Br\"{u}ll (HERMES), E. Hughes (E155), Z.-E. Meziani (E94-010), S.Rock (E155)
and A. Thomas (GDH).


\begin{thebibliography}{9}
\bibitem{smc_a1} SMC, B. Adeva et al., Phys. Rev. D58 (1998) 112001.    
\bibitem{e142_tot} E142, P.L. Anthony et al., Phys. Rev. D54 (1996) 6620.
\bibitem{e143_tot} E143, K. Abe et al., Phys. Rev. D58 (1998) 112003.    
\bibitem{e154_1} E154, K.Abe et al., Phys. Rev. Lett. 79 (1997) 26.
\bibitem{e154_2} E154, K.Abe et al., Phys. Lett. B404 (1997) 377.  
\bibitem{e154_3} E154, K.Abe et al., Phys. Lett. B405 (1997) 180.  
\bibitem{e155_1} E155, P.L. Anthony et al., hep-ex/9901006. 
\bibitem{e155_2} E155, P.L. Anthony et al., hep-ph/9902412. 
\bibitem{e155_pub} E155, P.L. Anthony et al., SLAC-PUB-8041 (March 1999).
\bibitem{hermes_1} HERMES, K.Ackerstaff et al., Phys. Lett. B404 (1997) 383.
\bibitem{hermes_2} HERMES, A.Airapetian et al., Phys. Lett. B442 (1998) 484.
\bibitem{hermes_3} HERMES, K.Ackerstaff et al., Phys. Lett. B444 (1998) 531.
\bibitem{smc_perp_p} SMC, D.Adams et al., Phys. Lett. B336 (1994) 125.
\bibitem{smc_perp_d} SMC, D.Adams et al., Phys. Lett. B396 (1997) 338.
\bibitem{song} X. Song, Phys. Rev. D54 (1996) 1955.
\bibitem{stratman} M. Stratmann, Z.Phys. C60 (1993) 763.
\bibitem{emc} EMC, J.Ashman et al., Phys. Lett. B206 (1998) 364.  
\bibitem{ej} J. Ellis and R.L. Jaffe, Phys. Rev. D9 (1974) 1444; 
{\it ibid} D10 (1974) 1669.
\bibitem{gdh_th} S.B. Gerasimov, Sov. J. Nucl. Phys. 2 (1966) 430;
S.B. Drell, A.C. Hearn, Phys. Rev. Lett. 16 (1966) 908.
\bibitem{dhg_protvino} H. Dutz, Proc. 13th Int. Symp. on High Energy Spin Physics,
Protvino (Sept. 1998).
\bibitem{dhg_mod} D.Drechsel et al., hep-ph/9810480-v2 (26 Jan 1999).
\bibitem{meziani_proc} Z.E. Meziani, Proc. of this workshop.      
\bibitem{ba_kw} B.Badelek and J. Kwiecinski, Phys. Lett. B295 (1992) 263.
\bibitem{smc_qcd} SMC, B. Adeva et al., Phys. Rev. D58 (1998) 112002.
\bibitem{smc_t15} SMC, B.Adeva et al., CERN-EP/99-61 (April 1999).
\bibitem{abfr} G.Altarelli et al., Acta Phys. Pol. B29 (1998) 1145, hep-ph/9803237;\\
 S.Forte et al., hep-ph/9808462; \\
G.Altarelli et al., Nucl.Phys. B534 (1998) 277.
\bibitem{lss} E.Leader et al., Phys.Rev.D58 (1998) 114028; E.leader et al., 
Phys.Lett. B445 (1998) 232.
\bibitem{mrst} A.D. Martin et al., hep-ph/9803445.
\bibitem{grv} M.Gluck et al., Z.Phys. C67 (1995) 433.
\bibitem{bj} J.D. Bjorken, Phys.Rev. 148(1966) 1467; {\it ibid.} D1 (1970) 1376.
\bibitem{pol_hera} "Physics with polarized protons at HERA", Desy-proceeding-1998-01,
(February 1998).
\bibitem{smc_si} SMC, B. Adeva et al., Phys. Lett. B420 (1998) 180.
\bibitem{gs} T. Gehrmann and W. J. Stirling , Phys. Rev. D53 (1996) 6100.
\bibitem{deflo} D. de Florian et al., Phys. Rev. D57 (1998) 5803.
\bibitem{compass} COMPASS proposal, CERN/SPSLC 96-14, SPSC/P297;
CERN/SPSLC 96-30 (May 1996).
\end{thebibliography}
\end {document}